\documentclass[twocolumn,showpacs,preprintnumbers,amsmath,amssymb]{revtex4}
\usepackage{graphicx}% Include figure files
\usepackage{dcolumn}% Align table columns on decimal point
\usepackage{bm}% bold math

\begin{document}

\preprint{}

\title{Two non-comoving stiff fluids in radial motion and spherical symmetry}% Force line breaks with \\

\author{Valentin Kostov}
 \altaffiliation[]{valentin@uchicago.edu}%Lines break automatically or can be forced with \\
\affiliation{The University of Chicago, Department of Physics, 5720 S. Ellis Ave, Chicago, IL 60637}

\date{\today}% It is always \today, today,

\begin{abstract}
The problem of two stiff fluids (energy density = pressure) moving radially in spherical symmetry is treated.
The metric ansatz is chosen spherically symmetric, conformally static with a multiplicative separation of variables. The first fluid is described mathematically via a massless scalar field. The coordinate system is chosen comoving with the second fluid which the separation of variables requires to be stiff too. The fluids are interacting only gravitationally and their energy momentum tensors are separately conserved. The Einstein equations are reduced to a single nonlinear ODE of second order which is shown to lead to an Abel ODE. A few particular exact solutions were found using a polynomial ansatz. The two non-comoving gravitational sources in the solutions can be interpreted either as scalar fields or stiff fluids. A complete analysis is performed on the range of parameters for which the stiff fluid interpretation is physically acceptable. General formulas are derived for the conformal vectors of the solutions. By making the second fluid vanish, a few single scalar field solutions are generated some of which appear to be new. All solutions considered in this paper have a time-like singularity at the origin (except the trivial FRW one) and are not asymptotically flat (except the static one with k=0).\end{abstract}
\maketitle

%\section{\label{sec:level1}First-level heading:\protect\\ The line
%break was forced \lowercase{via} \textbackslash\textbackslash}
\section{Introduction}

The most economical way to explain the plethora of current cosmological data is the $\Lambda$CDM model which contains dark matter. Its interactions with the baryonic matter is assumed so weak that it still evades a direct experimental detection. The various matter components in $\Lambda$CDM have the same four-velocity. That is expected if they originated from the same component in the past or if they have had enough interaction to equalize their expansion rates. The existence of a weakly interacting component opens the possibility for non-comoving motion. Such a scenario cannot be realized within the homogeneous and isotropic FRW metrics. Those are diagonal in the usual comoving coordinates and require a diagonal energy momentum tensor through the Einstein equations. Non-comoving components  described as perfect fluids would necessarily produce off-diagonal elements in the energy-momentum tensor and are incompatible with the assumed homogeneity and isotropy of the $\Lambda$CDM metric.

Spacetimes with less demanding symmetries, of which the simplest are the plane or the spherically symmetric ones, allow naturally for a non-comoving motion of their matter sources. The published exact solutions for two non-comoving fluids are scarce due to the prohibitive mathematical complexity. The case of plane symmetry and two stiff fluids in irrotational motion was solved in \cite{let}. Shear free two-fluid solutions were considered in \cite{fer} but the interaction between the fluids was not fixed on physical grounds, a problem common to many exact solutions. Two non-comoving dusts in spherically symmetric radial motion were treated in \cite{ha1, ha2} where, by imposing the existence of a Killing vector, the problem was reduced to numerical integration of two coupled non-linear ODE's.

Obtaining non-comoving exact solutions is important not only as toy cosmological models but also, from purely mathematical point of view, they provide examples of solving systems of coupled nonlinear differential equations for which general methods yet do not exist. Non-comoving fluids could also arise in studies of star evolution, for example neutron stars \cite{bay}. 

When the system of Einstein PDE's is intractable, the usual methods to reduce the complexity are separation of variables or imposing the existence of Killing or homothetic vector fields which convert the PDE's into ODE's. Stiff fluids(energy density = pressure) appear often in the literature since they allow for a simple mathematical description in terms of massless scalar fields \cite{tab}.

The present paper was inspired by \cite{sal} in which exact solutions were derived in the context of a modified GR theory. As the authors point out, their equations are mathematically equivalent in GR to a scalar field conveniently interacting with a perfect fluid in spherical symmetry. That suggested a similar GR solution is also possible for a stiff fluid (described by a scalar field) interacting only gravitationally with another fluid. 

Analogously to \cite{sal} the metric ansatz in the present paper is chosen spherically symmetric, conformally static with a multiplicative separation of variables. The first fluid is described by a massless scalar field. The coordinates are chosen comoving with the second fluid which the separation of variables require to be stiff too. The fluids are interacting only gravitationally and their energy momentum tensors are separately conserved. The Einstein equations are reduced to a single nonlinear ODE of second order which is shown to lead ultimately to an Abel ODE. A few particular polynomial solutions are found. The two non-comoving matter sources in the solutions can be interpreted either as scalar fields or as stiff fluids. A complete analysis is performed on the range of parameters for which the stiff fluid interpretation is physically acceptable. General formulas are derived for the conformal vectors of the solutions. By making the second fluid vanish, a few single scalar field solutions are generated some of which appear to be new. 

The units employed throughout this paper are $8\pi G = 1$ and the metric signature is (- + + +).

\section{Correspondence between a stiff fluid and a massless scalar field}

As was shown in \cite{tab} for any perfect fluid with a nonzero pressure,  the energy momentum tensor and its conservation can be written in terms of a scalar function. The stiff fluids enjoy  particularly simple equations that are mathematically equivalent to those of massless scalar fields as outlined below.  

The energy momentum tensor of a stiff fluid in the usual form is
\begin{equation}
	T_{\mu\nu}=2\epsilon \,  U_\mu U_\nu + \epsilon \,  g_{\mu \nu},
\end{equation}
where $\epsilon$ is the energy density (and pressure) and $U_\mu$ is the four-velocity. If the motion is irrotational, $U_\mu$ is proportional to a gradient of a scalar function $\Phi$, and normalized \cite{ste}:
\begin{equation}\label{4vel}
	U_\mu=\frac{\Phi_{,\mu}}{\sqrt{-\Phi_{,\lambda} \, \Phi^{,\lambda}}}\,\,,
\end{equation}
where comma denotes partial derivatives. The energy momentum conservation, $T^{\mu\nu}_{\;\;\;\; ; \mu}=0$, projected parallel and perpendicular to $U_\mu$ leads to the two equations \cite{ste}:
\begin{eqnarray}
	U^{\nu} \epsilon_{,\nu}&=&-2\, \epsilon \, U^\nu_{\;\; ;\nu},\label{cen}\\
	(g^\nu_\mu + U^\nu U_\mu) \epsilon_{, \nu} &=& -2 \epsilon U^\nu U_{\mu ; \nu}.\label{cmom}
\end{eqnarray}
One can always redefine  $\Phi \rightarrow F(\Phi)$ with an arbitrary function F , without changing $U_\mu$,  until  (\ref{cmom}) is satisfied identically \cite{tab}. Then the energy density is given by
\begin{equation}\label{eps}
	\epsilon=-\Phi_{,\lambda} \, \Phi^{,\lambda}
\end{equation}
and (\ref{cen}) leads to the equation
\begin{equation}
	\Phi^\nu_{;\nu} = \frac{1}{\sqrt{-g}}(\Phi_{,\mu} g^{\mu \nu} \sqrt{-g})_{,\nu}=0.
\end{equation}

Substituting (\ref{4vel}) and (\ref{eps}) in (1), obtains the energy momentum tensor
\begin{equation}
	T^{(n)}_{\mu\nu}= 2\, \Phi_{,\mu} \Phi_{,\nu} -\Phi_{,\lambda} \, \Phi^{,\lambda}\, g_{\mu\nu},
\end{equation}
which up to a factor of two is the usual tensor for a scalar field of zero mass and a vanishing potential.
The conservation equation of (7) is indeed (6). 

For fluids that are not stiff, the equations analogous to (6) and (7) are more complicated and not amenable to the separation of variables employed later in this paper.

\section{Metric and Einstein equations}
The most general diagonalized spherically symmetric and conformally static metric with multiplicative separation of variables is 

\begin{equation}
	ds^2=F(t)\left\{ \,-A^2(r) \, dt^2 + B^2(r) \, dr^2 + C^2(r)\, d\Omega^2 \,\right\},
\end{equation}
where $d\Omega^2=d\theta^2 + \sin^2(\theta) \, d\phi^2$. The above notation is similar to the one in \cite{sal} but here F(t) is not squared and is pulled in front as a conformal factor. That makes the equations for F(t) easily solvable. 

The total energy momentum tensor of the two components is 
\begin{equation}
	T_{\mu\nu} = T^{(n)}_{\mu\nu} + T^{(c)}_{\mu\nu},
\end{equation}
where $T^{(n)}_{\mu\nu}$ is given by (7) and represents a stiff fluid moving along the radial direction; $T^{(c)}_{\mu\nu}$ represents a perfect fluid which will necessarily turn out to be stiff too:
\begin{equation}
	T^{(c)}_{\mu\nu} = (\rho + p) V_\mu V_\nu + p \, g_{\mu \nu}.
\end{equation}
Just like in \cite{sal}, for simplicity, the coordinate system is chosen comoving to the four-velocity   
\begin{equation}
	V^\mu = (\frac{1}{F^{1/2}|A|}, 0, 0, 0)
\end{equation}
of the second fluid - the one denoted with "c"  for "comoving". Note that $V^\mu$ has zero shear.

There are four independent non-zero Einstein equations $G_{\mu\nu} \equiv R_{\mu\nu} - (R/2)g_{\mu\nu} = T_{\mu\nu}$:
\begin{widetext}
\begin{eqnarray}
	G_{tt}&=&\frac{3 \dot{F}^2}{4 F^2}+\frac{A^2}{B^2} \left(- \frac{2 C^{\prime \prime}}{C} + \frac{2 B^\prime C^\prime}{B C} - \frac{C^{\prime 2}}{C^2} + \frac{B^2}{C^2}\right)= %\nonumber \\
	\dot{\Phi}^2+ \frac{A^2}{B^2}\Phi^{\prime 2} + A^2 F \rho\\
	G_{tr}&=&\frac{A^\prime}{A} \frac{\dot{F}}{F} = 2 \Phi^\prime \dot{\Phi}\\
	G_{rr}&=&\frac{B^2}{A^2}\left\{-\frac{\ddot{F}}{F} + \frac{3 \dot{F}^2}{4 F^2}+\frac{A^2}{B^2} \left(\frac{A^\prime C^\prime}{A \, C} + \frac{C^{\prime 2}}{C^2} - \frac{B^2}{C^2}\right) \right\}=%\nonumber\\
	\frac{B^2}{A^2}\left(\dot{\Phi}^2+\frac{A^2}{B^2}\Phi^{\prime 2}+A^2 F p \right)\\
	G_{\theta\theta}&=&\frac{C^2}{A^2} \left\{ -\frac{\ddot{F}}{F} + \frac{3 \dot{F}^2}{4 F^2} +\frac{A^2}{B^2} \left( \frac{C^{\prime\prime}}{C}+\frac{A^{\prime\prime}}{A} + \frac{A^\prime C^\prime}{A \, C} - \frac{B^\prime C^\prime}{B C} - \frac{A^\prime B^\prime}{A B}\right)\right\}=% \nonumber\\
	\frac{C^2}{A^2} \left(\dot{\Phi}^2 - \frac{A^2}{B^2}\Phi^{\prime 2} +A^2 F p \right),
\end{eqnarray}
\end{widetext}
with dot denoting $\partial/\partial t$ and prime denoting $\partial/\partial r$. As usual in the separation of variables method, the simplifying equation is the off-diagonal one (13). Following \cite{sal}, an additive separation of variables is chosen for the scalar field
\begin{equation}
	\Phi(r,t)=W(r)+T(t).
\end{equation}
Substituting that in (13) and solving for the scalar field functions, W and T, in terms of the metric functions gives
\begin{eqnarray}
	T(t)=\frac{T_0}{2} \ln \left(\frac{F(t)}{F_c}\right)\\
	W(r)=\frac{1}{T_0} \ln \left(\frac{A(r)}{A_c}\right),
\end{eqnarray}
where $T_0, F_c,$ and $A_c$ are constants. The values of $A_c$ and $F_c$ are of no importance since they do not influence the derivatives of $\Phi$ which appear in the Einstein equations. 

Note that at the boundary points of the $r$ and $t$ intervals on which the metric is defined, $A(r)$ or $F(t)$ vanish and the scalar field $\Phi$ becomes infinite. Vanishing of $A(r)$ signifies a timelike singularity, vanishing of $F(t)$ - a spacelike (cosmological) singularity.

\section{Separate energy momentum conservation}\label{encons}
The Einstein equations are incomplete without specifying the energy transfer between the two fluids. The simplest scenario is to assume they interact only gravitationally i.e. their energy momentum tensors are divergenless. The conservation equation for the comoving fluid, $T^{(c) \mu\nu}_{\,\,\,\,\,\,\,\,\,\,\,\,\,\, ;\nu}=0$ gives the following two equations
\begin{eqnarray}
	\frac{\dot{\rho}}{\rho} + \frac{3 \dot{F}}{2 F} \left(1+ \frac{p}{\rho} \right) = 0 \\
	\frac{p^{\,\prime}}{\rho} + \frac{A^\prime}{A}\left(1+ \frac{p}{\rho} \right) = 0. 
\end{eqnarray}
Assuming a barotropic equation of state, $p=\alpha \rho, \,\, \alpha=$const, the solution of (19) and (20) is
\begin{equation}
	\rho(r,t)=\frac{\rho_0}{F(t)^{3(1+\alpha)/2} A(r)^{(1+\alpha)/\alpha}},
\end{equation}
where $\rho_0$ is an integration constant. The various terms in (12) depend either on $r$ or on $t$. The same should apply to the term $A^2 F \rho$ to enable the separation of variables. The term is a function of $t$ only if A(r)=const or $\alpha=1$ (stiff fluid) and it is a function of $r$ only if F=const or $\alpha=-1/3$. In the case of A=const, eq. (18) leads to $\Phi^\prime=0$ which by eq. (2) implies $U^r=0$ i.e. the two fluids are both comoving with the coordinate system - that case will be considered further in section \ref{kzero}. The case of F(t)=const, using eq. (17), leads to $\dot{\Phi}=0$ which by eq. (2) implies the four-velocity U has a zero time component which is not physical. The case $\alpha = -1/3$ does not describe a known type of matter. The only possibility remaining is that the second fluid is also stiff with an energy density of
\begin{equation}
	\rho(r,t)=p(r,t)=\frac{\rho_0}{F(t)^3 A(r)^2}.
\end{equation}

The conservation equation of the first non-comoving stiff fluid, $T^{(n) \mu\nu}_{\,\,\,\,\,\,\,\,\,\,\,\,\,\, ;\nu}=0$, leads to the equation
\begin{equation}
	\left(C^2 \frac{A}{B} W^\prime \right)^\prime \, \frac{A}{B C^2} = \frac{1}{F} (\dot{T} F)^\cdot.
\end{equation}
Substituting (17) and (18) in (23) gives
\begin{equation}
	\left(C^2 \frac{A^\prime}{B} \right)^\prime \, \frac{A}{B C^2} = \frac{T_0^2 \ddot{F}}{2 F} = \frac{T_0^2}{2} k,
\end{equation}
where 
\begin{equation}
	k\equiv \ddot{F}/F
\end{equation}
is a separation of variables constant. 

The Einstein equations guarantee the conservation of the total energy momentum tensor. Therefore, if one of the tensors $T^{(c) \mu\nu}$ and $T^{(n) \mu\nu}$ is conserved, so is the other. Thus the Einstein equations and (22) are sufficient to specify the system; eq. (24) is their consequence. Nevertheless, it can and will be used later to simplify the analysis.

\section{ Equivalent system of equations}

The Einstein equations that need to be satisfied are (12), (14), and (15) in which (17), (18) and (22) were substituted. A more convenient system of equations, containing fewer terms, is obtained by forming independent linear combinations of those:

\begin{widetext}
\begin{eqnarray}
	\frac{A^2}{C^2}\left( \left( \frac{(C^2 A^2)^\prime}{A B}\right)^\prime \frac{1}{A B} - 2 \right) = 4\frac{\ddot{F}}{F}+ \frac{\dot{F}^2}{F^2}(T_0^2-3) + 4\frac{\rho_0}{F^2} = k (T_0^2 +1) \\
	\frac{A^2}{B^2}\left( 2 \frac{A^\prime C^\prime}{A C} + \frac{C^{\prime 2}}{C^2} - \frac{B^2}{C^2}\right) - \frac{A^{\prime 2}}{B^2 T_0^2} = \frac{\ddot{F}}{F}+ \frac{\dot{F}^2}{4 F^2}(T_0^2-3) + \frac{\rho_0}{F^2} = \frac{k}{4} (T_0^2 +1) \\
	\left( \frac{C^2 (A^2)^\prime}{A B} \right)^\prime \frac{A^2}{(AB) C^2}=3\frac{\ddot{F}}{F}+ \frac{\dot{F}^2}{F^2}(T_0^2-3) + 4\frac{\rho_0}{F^2} = k T_0^2 
\end{eqnarray}
\end{widetext}
The first equation above corresponds to $2 (G_{rr} A^2/B^2 + G_{\theta\theta} A^2/C^2)$, the second to $G_{rr}A^2/B^2$, the third to $2 (G_{tt}+G_{rr} A^2/B^2 + 2 G_{\theta\theta} A^2/C^2)$, all separated in variables. The rightmost sides of the equations are the separation constants determined in the following way. The left hand side of (28) is just twice the conservation equation(24). As mentioned before, (24) is redundant and should follow automatically from (26) - (28). Nevertheless, it can be used as a shortcut implying that the right hand side of (28) is $k T_0^2$. Another consequence  of (24) is that $\ddot{F}/F = k$. That allows to solve (28) for the expression:
\begin{equation}
	\frac{\dot{F}^2}{F^2}(T_0^2-3) + 4\frac{\rho_0}{F^2}=k(T_0^2-3).
\end{equation}
Substituting that in (26) and (27) produces their rightmost sides.

\section{Solution of the time equations}
The principal time equation is (25) with the solution
\begin{equation}
	F(x)=\left\{     
	\begin{array}{lr}   
	f_1 \exp(\sqrt{k} \,\, t) + f_2 \exp(- \sqrt{k}\,\,  t), \,\, & k>0\\       
	f_1 t + f_2, \,\, & k=0 \\
	f_1 \sin(\sqrt{-k} \,\, t + f_2), \,\, & k<0,     
	\end{array}   
	\right.
\end{equation}
where $f_1$ and $f_2$ are constants. In order to satisfy the rightmost sides of (26)-(28), F(t) must satisfy (29). Substituting (30) in (29) determines the required values of $\rho_0$:
\begin{equation}
	\rho_0=\left\{     
	\begin{array}{lr}   
	k(T_0^2-3)f_1 f_2, \,\, & k>0\\       
	-(T_0^2-3)f_1^2/4, \,\, & k=0 \\
	k(T_0^2-3)f_1^2/4, \,\, & k<0.     
	\end{array}   
	\right.
\end{equation}

\section{Radial gauge and radial equations}\label{gauge}
A coordinate transformation involving only the radial coordinate, $r=r(\tilde{r})$, induces the following transformation in the metric functions
\begin{equation}
	\begin{array}{lr}
	A(r) \rightarrow \tilde{A}(\tilde{r}) = A(r(\tilde{r}))\\
	B(r) \rightarrow \tilde{B}(\tilde{r}) = B(r(\tilde{r})) dr/d\tilde{r}\\
	C(r) \rightarrow \tilde{C}(\tilde{r}) = C(r(\tilde{r})).
	\end{array}
\end{equation}
The coordinate freedom in $r$ can be used to select a coordinate system in which the Einstein equations have a simple form. Many choices are possible but a glance at (26) and (28) suggests $\tilde{A} \tilde{B} =1$ in the new coordinates. The same gauge was used in \cite{sal}. From (32), the required transformation, $r(\tilde{r})$, is obtained by solving the differential equation
\begin{equation}
	A(r) B(r) \frac{dr}{d\tilde{r}} = 1
\end{equation}
for $\tilde{r}(r)$ and inverting that to $r(\tilde{r})$. The integration of (33) and the inversion are always possible since A and B, being metric functions, should be nonzero. The solution $\tilde{r}(r)$ is fixed up to an integration constant implying that all radial coordinates for which $\tilde{A} \tilde{B} =1$ are related by translations: $\tilde{r}^* = \tilde{r} + const$. That residual freedom in $\tilde{r}$ will be apparent in the Einstein equations later and can be used to absorb constants. From now on, it will be assumed that the necessary coordinate transformation has already been effected and $AB=1$ in the new coordinates (dropping the tildes for convenience). In this gauge, the radial equations (26)-(28) simplify to
\begin{eqnarray}
		\frac{A^2}{C^2}\left( (C^2 A^2)^{\prime \prime} - 2 \right) = k (T_0^2 +1) \\
	  A^4\left( 2 \frac{A^\prime C^\prime}{A C} + \frac{C^{\prime 2}}{C^2} - \frac{1}{A^2 C^2} - \frac{A^{\prime 2}}{A^2 T_0^2}\right) = \frac{k}{4} (T_0^2 +1) \\
	\left( C^2 (A^2)^\prime \right)^\prime \frac{A^2}{C^2}=k T_0^2.
\end{eqnarray}

\section{Radial solution for \lowercase{k=0}} \label{kzero}
When $k=0$, (34) implies
\begin{equation}
	A^2 C^2 = (r-r_1)(r-r_2) \equiv R(r),
\end{equation}
where $r_1$ and $r_2$ are integration constants. Substituting $C^2=R/A^2$  and $k=0$ in (36) and integrating twice gives
\begin{equation}
	A^2 = A_0^2 \exp \left(\alpha \int \frac{dr}{R(r)}\right),
\end{equation}
where $\alpha$ and $A_0^2$ are constants. Now $C$ can be found:
\begin{equation}
	C^2 = \frac{R(r)}{A_0^2} \exp \left(-\alpha \int \frac{dr}{R(r)}\right).
\end{equation}
The last equation to be satisfied, (35), will put restrictions on $\alpha, r_1,$ and $r_2$. Substituting (38) and (39) in (35) leads to
\begin{equation}
	R^{\, \prime \,2} - 4\, R - \alpha^2 (1 + 1/T_0^2) =0.
\end{equation}
Inserting the explicit form of $R(r)$ in the above gives the restriction
\begin{equation}
	\alpha = \pm \frac{r_2-r_1}{\sqrt{1+1/T_0^2}}.
\end{equation}
An analogous solution was obtained in \cite{sal}. The constant $\alpha$ must be real since it enters the derivative $A^\prime/A = \alpha/(2 R)$. This is possible only if $r_1$ and $r_2$ in (41) have equal imaginary parts. The roots $r_1$ and $r_2$ must be complex conjugate to each other to produce the real-valued polynomial $R(r)$. The last two statements are compatible only when $r_1$ and $r_2$ are real. The two distinct cases, $r_1=r_2$ and $r_1 \ne r_2$ are now discussed.

The case $A=const$ of two comoving fluids was mentioned in section \ref{encons}. It implies $k=0$ from (36), $\alpha=0$ from (38) and $r_1=r_2$ from (41). The metric is spatially flat FRW with Big Bang or Big Crunch depending on the sign of $T_0$ as will be discussed later. The energy density (5) is non-negative if $A=const$. The energy density (22) of the other fluid is non-negative if $T_0^2\le 3$, see (31). 

In the case of $r_1\ne r_2$ the two fluids are not comoving with each other. The metric functions become
\begin{eqnarray}
	A^2(r)&=&A_0^2 \left| \frac{r-r_2}{r-r_1} \right| ^n\\
	C^2(r)&=&\frac{1}{A_0^2} (r-r_1)(r-r_2)\left| \frac{r-r_2}{r-r_1} \right| ^{-n}, \nonumber
\end{eqnarray}
where the power $n= \pm (1+1\,/ \,T_0^2\, )^{-1/2}$. The sign of $n$ is the sign in (41) chosen  for  $\alpha$. 

\section{Particular radial solutions for \lowercase{ $ k\ne 0$}}
For nonzero $k$, the radial equations (34)-(36) can be reduced to a single equation involving the function
\begin{equation}
	S(r) \equiv A^2 C^2, \,\, S^{\prime \prime} \ne 2.
\end{equation}
The heuristics behind taking that combination is trying to imitate the way the $k=0$ solution was obtained as far as possible. A big advantage of that choice is that equations (34) and (43) can be solved $\it{algebraically}$ for $A^2$ and $C^2$ in terms of $S$:
\begin{eqnarray}
	A^2 &=& \sqrt{\frac{k(T_0^2 +1) S}{S^{\prime \prime} -2}}\\
	C^2 &=& \sqrt{\frac{S(S^{\prime \prime} -2)}{k(T_0^2 +1)}},\nonumber
\end{eqnarray}
where $S>0$ because of (43) and $k/(S^{\prime \prime} -2)>0$ because of (34). Substituting the above in (36) obtains
\begin{equation}
	\left( S^{\,\prime} - \frac{S S^{\prime \prime \prime}}{S^{\prime \prime} -2}\right)^\prime = \frac{2 T_0^2}{T_0^2 + 1} (S^{\prime \prime} -2).
\end{equation}
Integrating that equation once gives
\begin{equation}
	(T_0^2+1)S S^{\prime \prime \prime}=(S^{\prime \prime} -2)(S^{\,\prime} (1-T_0^2) + 4 \,r \,T_0^2 + C_1),
\end{equation}
where $C_1$ is the integration constant. Equation (46) can be integrated again resulting in
\begin{eqnarray}
	(T_0^2+1)S S^{\prime \prime} - (S^{\,\prime})^2 - S^{\,\prime}(4T_0^2r + C_1) + 2(T_0^2 +1)S \,\,\,\,\\
	 + (4T_0^2 r^2 + 2 C_1 r + C_2)=0,\nonumber
\end{eqnarray}
where $C_2$ is the new integration constant. The equation left to be satisfied is (35) - it will put restrictions on (47). Substituting (44) in (35) gives
\begin{eqnarray}
	-(T_0^2+1)\left(\frac{ S^{\prime \prime \prime} S}{S^{\prime \prime}-2}\right)^2 + 2(T_0^2+1)S^{\,\prime} \left(\frac{ S^{\prime \prime \prime } S}{S^{\prime \prime}-2}\right) \\
	+ (3 T_0^2 -1)S^{\,\prime \,2} - 4T_0^2(S^{\prime \prime}+2)S =0. \nonumber
\end{eqnarray}
The order of this equation can be reduced by solving (46) for $S^{\prime \prime \prime} S/(S^{\prime \prime}-2)$ and substituting that back in (48). The result is 
\begin{eqnarray}
	(T_0^2+1)SS^{\prime \prime} - S^{\,\prime \,2} - S^{\,\prime}(4 T_0^2 \, r + C_1) + 2(T_0^2 +1)S \,\,\,\,\\
	+ \frac{1}{4T_0^2}(4 T_0^2 \, r + C_1)^2 =0.\nonumber
\end{eqnarray}
Comparing that to (47) reveals that the restriction put by (35) on (47) is simply $C_2=C_1^2/(4 T_0^2)$.

In the above treatment, the Einstein equations (34)-(36) were reduced to (49) - a single non-linear ODE of second order. The integration constant $C_1$ in (49) expresses the translational freedom of the radial coordinate discussed in section \ref{gauge}: if S(r) is a solution of (49) for a given constant $C_1$, then $\bar{S}(r) = S(r-r_0), r_0 = const$ is also a solution of (49) but for another constant $\bar{C_1} = C_1 - 4 T_0^2 r_0$.

Three particular solutions of (49) were found by trying a polynomial ansatz for $S(r)$. These are:
\begin{eqnarray}
	S(r) &=& \frac{2 T_0^2}{T_0^2 -1}(r-r_0)^2 \\
	A^2(r) &=& \sqrt{k}\,\,|T_0(r-r_0)| \nonumber\\
	C^2(r) &=& \frac{2}{\sqrt{k}}\left| \frac{T_0}{T_0^2-1}(r-r_0)\right|, \nonumber
\end{eqnarray}
where $T_0^2>1, \,\, r\ne r_0, \,\, k>0$;

\begin{eqnarray}
	S(r) &=& d-(r-r_0)^2 \\
	A^2(r) &=& \sqrt{(-k/3)(d-(r-r_0)^2)} \nonumber\\
	C^2(r) &=& \sqrt{(-3/k)(d-(r-r_0)^2)}, \nonumber
\end{eqnarray}
where $T_0^2=1/3,\,\, d>0, \,\,k<0, \,\,|r-r_0|<\sqrt{d}$;

\begin{eqnarray}
	S(r) &=& a(r-r_0)^4+(r-r_0)^2 + b(r-r_0) \\
	A^2(r) &=& \frac{1}{3}\sqrt{\frac{k}{a} \left( a(r-r_0)^2 + 1 + \frac{b}{r-r_0} \right)} \nonumber\\
	C^2(r) &=& 3\sqrt{\frac{a}{k} \left( a(r-r_0)^4+(r-r_0)^2 + b(r-r_0) \right)} , \nonumber
\end{eqnarray}
where $T_0^2=1/3,\,\, a\ne 0, \,\, S(r)>0, \,\, k/a>0$.\\

In the above solutions, $a, b, d$ are constants and $r_0$ is a constant reflecting the translational freedom in $r$. The constraints after each solution come from the requirements $S>0$ and $k/(S^{\prime \prime} -2)>0$ mentioned after equation (44) and from requiring positivity of the expressions inside the square roots . Since the solution (52) depends on two constants, $a$ and $b$, it captures the $a\ne 0$ part of a general solution for $T_0^2=1/3$.

Although the main equation (49) is not particularly complicated in appearance, its general solution for arbitrary $T_0$ could not be found. The following sequence of variable transformations: $C_1=0$ (by translation in $r$); $S(r)=u(r)r^2$; $r=\exp(x)$; reduce (49) to an equation for $u(x)$ in which the independent variable $x$ does not appear explicitly:
\begin{eqnarray}
	(T_0^2+1)u u_{xx} + (3T_0^2-1)(u_x -2)u- u_x^2 - 4 T_0^2 u_x \nonumber \\
	+ 2(T_0^2-1)u^2 + 4 T_0^2=0, \nonumber
\end{eqnarray}
where subscripts denote ordinary derivatives.
The standard technique for solving such type of equations is the substitution for the first derivative $u_x=q(u)$ (hence $u_{xx} = q(u) q_u$) which converts the above equation to an Abel type ODE for $q(u)$:
\begin{eqnarray}
	qq_u - \frac{q^2}{u(T_0^2+1)}+\frac{(-u-4 T_0^2+3 To^2 u)\,q}{u(T_0^2+1)} \nonumber\\
	 +\frac{2(u-1)(-u+T_0^2 u-2 T_0^2)}{u (T_0^2+1)} = 0, \nonumber
\end{eqnarray}
for which general methods of solving are not known at present. Two particular solutions of (49) are known, which translate into particular solutions of the above Abel equation. These are solution (50) and the following
$$S(r) = (r-r_0)^2 + b(r-r_0) + \frac{b^2}{4(T_0^2 +1)}.$$
The above $S(r)$ was not mentioned before since it has the forbidden property $S^{\prime \prime} =2$ and thus it cannot generate solutions for $A^2$ and $C^2$. For Riccati types of ODE's, which are particular cases of Abel,  there are techniques to obtain general solutions starting from one or more particular solutions \cite{mur}. Unfortunately, the equation here is of the more general Abel type for which such methods are not currently available.

\section{Physicality of the stiff fluid interpretation}

In any solution, the metric functions $F, A^2,$ and $C^2$ are required to be positive to produce the right metric signature. Additionally, if the two matter sources are interpreted as stiff fluids, their energy densities must be non-negative and the four-velocities must be future oriented. The energy density of the comoving fluid is non-negative when $\rho_0 \ge 0$. The energy density (5) of the first fluid written more explicitly is
\begin{equation}
	\epsilon = \frac{1}{4 F A^2} \left\{ \left(T_0 \frac{\dot{F}}{F}\right)^2 - \left(\frac{(A^2)^\prime}{T_0}\right)^2\right\}
\end{equation}
and its four-velocity (2) is 
\begin{equation}
	U^\mu = \frac{1}{2 F \sqrt{\epsilon}}\left( -\frac{T_0 \dot{F}}{A^2 F}, \,\, \frac{(A^2)^\prime}{T_0},\,\, 0,\,\, 0\right).
\end{equation}
The four-velocity is future oriented, $U^{\,t}>0$, when $T_0 \dot{F}<0$. Hence the coordinate system must be expanding, $\dot{F}>0$, when $T_0<0$ and contracting, $\dot{F}<0$, when $T_0>0$. The allowed range of the time coordinate is restricted by the inequalities $F>0$, $\dot{F}<0 \,\, \text{or}\,\, >0$ (depending on the sign of $T_0$) and the non-negativity of $\epsilon$:
\begin{equation}
	\left|\frac{\dot{F}}{F}\right| \ge \frac{|(A^2)^\prime|}{T_0^2}.
\end{equation}
The allowed intervals for $r$ are determined by the conditions $A^2>0, \,\, C^2>0$ and are bounded by points at which $A^2$ or $C^2$ become zero or infinite.

The $k=0$ solutions in (42) have a time function $F(t) = f_1 t + f_2$. If $T_0<0$ (expansion), the allowed time interval is $t>-f_2/f_1, \,\, f_1>0$ to satisfy the conditions $F>0, \,\, \dot{F}>0$. At time $t \rightarrow -f_2/f_1$ the energy densities (22) and (53) become infinite signifying a spacelike singularity, a Big Bang. Note that the velocity component $U^{\,r}$ can be negative if the first fluid is expanding slower than the one comoving with the coordinates. Analogously, the $T_0>0$ case (contraction) describes a spacetime ending with a Big Crunch at $t \rightarrow -f_2/f_1$ and the time is restricted to $t<-f_2/f_1, \,\, f_1<0$. If $r_2$ denotes the larger root in (42), positivity of the metric function $C^2$ requires $r<r_1$ or $r>r_2$. Without loss of generality, one can take for the allowed range $r>r_2$. A simple translation in $r$ can set $r_2$ to the conventional zero value for the origin. Since $|n|<1$, $C^2(r)$ does not become infinite even at $r_1$ or $r_2$. If $n>0 \, (n<0)$ the function $A^2(r)$ vanishes at $r=r_2\, (r=r_1)$ which makes the energy densities (22) and (53) infinite, signifying a timelike singularity at $r_2(r_1)$. The energy density (22) is non-negative if $T_0^2\le 3$,  see (31). The hardest physical requirement to satisfy is the non-negativity of $\epsilon$, (55). Written explicitly, this is
\begin{equation}
	\left| t + \frac{f_2}{f_1}\right| \le \frac{\sqrt{T_0^2(T_0^2 + 1)}}{A_0^2 |r_2-r_1|}|r-r_1|^{1+n} |r-r_2|^{1-n}.
\end{equation}
Since $|n|<1$ and $r_1 \ne r_2$ in the non-comoving case, the above inequality restricts $r$ for fixed $t$ or restricts $t$ for fixed $r$. Thus the energy density $\epsilon$ is not non-negative over the full allowed ranges of the coordinates. A similar problem occurs sometimes for a single stiff fluid in plane symmetry \cite{tab}. There the energy momentum tensor was reinterpreted as an anisotropic fluid in regions of negative $\epsilon$. If the non-comoving component in the present article is interpreted simply as a scalar field instead of a stiff fluid, the problem of negative $\epsilon$ becomes irrelevant.

The $k>0$ solution (50) has a time function $F=f_1 \exp (\sqrt{k}\,\,t) + f_2 \exp (-\sqrt{k}\,\,t)$. When the coordinates are expanding $(T_0<-1<0)$, the conditions $F>0$ (metric function) and $\dot{F}>0$ (future oriented four-velocity) lead to the restrictions
\begin{equation}
	f_1>0, \,\, t>t_* = \frac{1}{2 \sqrt{k}} \ln \left|\frac{f_2}{f_1}\right|.
\end{equation}
The corresponding restrictions for contracting coordinates $(T_0>1>0)$ are
\begin{equation}
	f_2>0, \,\, t<t_* = \frac{1}{2 \sqrt{k}} \ln \left|\frac{f_2}{f_1}\right|.
\end{equation}
The non-negativity of $\epsilon$ condition (55), written explicitly is
\begin{equation}
	\left|\frac{f_1 \exp (\sqrt{k}\,\,t) - f_2 \exp (-\sqrt{k}\,\,t)}{f_1 \exp (\sqrt{k}\,\,t) + f_2 \exp (-\sqrt{k}\,\,t)} \right| \ge \frac{1}{|T_0|}.
\end{equation}
For expanding coordinates $(T_0<-1, \,\, f_1>0)$, (59) translates into 
\begin{equation}
	\exp(2\sqrt{k} \,\, t) \ge \frac{|T_0|+1}{|T_0|-1} \,\cdot \, \frac{f_2}{f_1}.
\end{equation}
If $f_2\le 0$, (60) is not restrictive at all and the final allowed range of $t$ is given by (57); the spacetime starts with a Big Bang at $t_*$. If $f_2 >0$, (60) is more restrictive than (57) and the allowed range of $t$ is given by
\begin{equation}
	t>\frac{1}{2 \sqrt{k}} \ln \left(\frac{|T_0|+1}{|T_0|-1} \,\cdot \, \frac{f_2}{f_1}\right)> t_*.
\end{equation}
In this case the stiff fluid interpretation fails at times too close to the Big Bang time $t_*$. The case of contracting coordinates $(T_0>1, \,\, f_2>0)$ has a similar analysis resulting in the following allowed ranges:
\begin{eqnarray}
	t&<&t_*, \,\,f_1<0\\
	t&<& \frac{1}{2 \sqrt{k}} \ln \left(\frac{T_0-1}{T_0+1} \,\cdot \, \frac{f_2}{f_1}\right)< t_*, \,\, f_1>0.\nonumber
\end{eqnarray}
When $f_1>0$ the stiff fluid interpretation fails at times too close to the Big Crunch time $t_*$. The metric functions in solution (50) vanish at $r=r_0$. Without loss of generality, the allowed range can be taken as $r>r_0$. A translation of the $r$ coordinate can always set $r_0$ to the conventional zero value for the origin.

The $k<0$ solution (51) does not allow for a stiff fluid interpretation over the full range of $t$ set by the $F>0$ condition. Similarly to the $k=0$ solution (42) discussed before, the energy density $\epsilon$ becomes negative when $r$ approaches the roots $r_0 \pm \sqrt{d}$. The problem stems from $(A^2)^\prime$ becoming infinite at the roots. The detailed analysis is left to the reader.

Solution (52) allows for a stiff fluid interpretation when the derivative
\begin{equation}
	(A^2)^\prime = \frac{1}{3} \sqrt{\frac{k}{a}} \frac{a(r-r_0) - b/(2(r-r_0)^2)}{\sqrt{a(r-r_0)^2 +1 + b/(r-r_0)}}
\end{equation}
remains finite at the end points of the allowed $r$ intervals. The end points are roots of $S(r)$ which are also zeros of $C^2(r)$. The denominator in (63) vanishes at any root of $S(r)$ except $r_0$. The only way for (63) to remain finite at such a root, $r_* \ne r_0$, is that its numerator also vanishes leading to the system of equations
\begin{eqnarray}
	a(r_* -r_0) - \frac{b}{2(r_*-r_0)^2}=0\\
	a(r_*-r_0) + 1 + \frac{b}{r_*-r_0}=0. \nonumber
\end{eqnarray}
The solutions of the system are $b=\pm2/(3\sqrt{-3a}), \,\, r_* = r_0 \mp 1/\sqrt{-3 a}$, where $a<0$ for real values. Unfortunately, $S(r)<0$ in the neighborhood of $r_*$ hence $r_*$ cannot be an edge point of an allowed $r$ interval and is of no further interest. The root $r_0$ has a different behavior. The derivative (63) is finite at $r=r_0$ only when $b=0$. If $b=0$ and $a<0$ then (63) is finite at $r_0$ but infinite at the other two roots $r_0 \pm 1/\sqrt{-a}$ and the stiff fluid interpretation fails close enough to the roots.

The case $b=0$ and $a>0 \,\,(k>0)$ is more benign since then $S(r)$ is positive everywhere except at the double root $r=r_0$. The allowed $r$ interval can be taken as $r>r_0$. The derivative (63) is zero at $r_0$ and increases asymptotically towards the value $\sqrt{k}/3$. The non-negativity condition (55) will be satisfied over the whole allowed $r$ interval if $|\dot{F}/F|\ge 3|(A^2)^\prime|_{\text{max}}= \sqrt{k}$:
\begin{equation}
	\left|\frac{f_1 \exp (\sqrt{k}\,\,t) - f_2 \exp (-\sqrt{k}\,\,t)}{f_1 \exp (\sqrt{k}\,\,t) + f_2 \exp (-\sqrt{k}\,\,t)} \right| \ge 1.
\end{equation}
When $T_0=-1/\sqrt{3}<0$ (expansion), the conditions $F>0, \, \dot{F}>0$ again lead to the restrictions (57). Taking into account $F>0, \, \dot{F}>0$ to open the absolute value in (65) leads to $f_2\le0$ so the final result is
\begin{equation}
	T_0=-1/\sqrt{3}, \,\, f_1>0, \,\, f_2 \le 0, \,\, t>t_*.
\end{equation}
This spacetime has expanding coordinates and a comoving fluid that starts with a Big Bang at $t_*$.

In the opposite case $T_0=+1/\sqrt{3}>0$ (contraction), the conditions $F>0, \, \dot{F}<0$ again lead to the restrictions (58). Using $F>0, \, \dot{F}<0$ to open the absolute value in (65) leads to $f_1\le0$ and the final restriction on $t$ is 
\begin{equation}
	T_0=+1/\sqrt{3}, \,\, f_1\le0, \,\, f_2 > 0, \,\, t<t_*. 
\end{equation}
This describes a spacetime with contracting coordinates and a comoving fluid that ends in a Big Crunch at $t_*$.

\section{Conformal vector fields}
It is important to know if the spacetimes obtained posses Killing, homothetic or more generally conformal vectors. By definition, a conformal vector field, $\xi^\mu$, satisfies
\begin{equation}
	\mathfrak{L}_\xi  \,\,g_{\mu\nu}= \psi(x) \,\, g_{\mu\nu}
\end{equation}
where $\mathfrak{L}$ is the Lie derivative along $\xi$ and $\psi$ is a scalar function. Here $g_{\mu\nu}$ is the metric (8) in the $B=1/A$ gauge. For simplicity, the conformal vector is restricted to the form $\xi^\mu = (a, b, m, n)$ where its components are functions of $(t,r)$ only. The off-diagonal components of (68) give $m=\text{const}, \,\, n=\text{const}$ and 
\begin{equation}
	\dot{b}=A^4 a^\prime.
\end{equation}
The diagonal components of (68) lead to $m=0$ and 
\begin{eqnarray}
	b^\prime &=& b \left( \frac{A^\prime}{A}+ \frac{C^\prime}{C}\right)\\
	\dot{a} &=& b \left( \frac{C^\prime}{C}- \frac{A^\prime}{A}\right)\\
	\psi(r,t)&=& a(t,r) \frac{\dot{F}}{F}+2 b(t,r)\frac{C^\prime}{C}.
\end{eqnarray}
Equation (70) can be integrated leading to the equivalent set of equations
\begin{eqnarray}
	b&=& f(t) AC\\
	\dot{a} &=& f(t) (AC^\prime - CA^\prime)\\
	a^\prime &=& \frac{C}{A^3}\dot{f}\\
	\psi(r,t)&=& a(t,r) \frac{\dot{F}}{F}+2 b(t,r)\frac{C^\prime}{C},
\end{eqnarray}
where $f(t)$ is an arbitrary function of time.

The solutions of (73)-(76) split into two classes. In the case of $f(t)=0$, the solution is 
\begin{equation}
	\xi^\mu = (a, 0, 0, n), \,\, \psi = a\dot{F}/F,
\end{equation}
where $a$ and $n$ are arbitrary constants. 

Additional solutions for $f(t) \ne 0$ are possible only if the consistency condition $(\dot{a})^\prime = (a^\prime)\dot{}\,\,$:
\begin{equation}
	\frac{A^3}{C}(AC^{\prime \prime} - CA^{\prime \prime})= \frac{\ddot{f}}{f} = const
\end{equation}
is satisfied. That condition (with const=0) is true for solution (42)  with $r_1=r_2$ (the FRW case). For solutions (50) and (51), the combination $AC^{\prime \prime} - CA^{\prime \prime}$ vanishes too since $C\propto A$. In all those cases $AC^{\prime \prime} - CA^{\prime \prime} = 0$ and then the extra solution for $\xi^\mu$ is 
\begin{eqnarray}
	a(t,r) &=& (AC^\prime - CA^\prime)\int f(t) dt + f^\prime(t) \int \frac{C}{A^3}dr\\
	b(t,r) &=& f(t) AC\nonumber\\
	m(t,r) &=& 0\nonumber\\
	n(t,r) &=& const\nonumber\\
	\psi(r,t)&=& a(t,r) \frac{\dot{F}}{F}+2 b(t,r)\frac{C^\prime}{C}\nonumber
\end{eqnarray}
where $f(t)$ is an arbitrary linear function, $f^\prime(t) = \text{const}$. Note that the above conformal vector applies only when $AC^\prime - CA^\prime$=const.

\section{Single scalar field solutions}
By a proper choice of the parameters, the energy density (31) of the comoving fluid can be set to zero leaving a solution with a single non-comoving stiff fluid which can be interpreted as a massless scalar field. The case of $k=0, \,\, T_0^2=3, \,\,n=\pm \sqrt{3}/2$ is the non-static scalar field solution described in \cite{hus}. The case of $k=0, \,\, f_1=0, \,\, T_0$ arbitrary and $n$ arbitrary gives a particular class of asymptotically flat static solutions published in \cite{agn}. 

Other possibilities for generating a single scalar field solution are $k>0, \,\, f_1 f_2=0$; $k>0, \,\, T_0^2=3$; $k<0, \,\, T_0^2=3$. These do not appear to have been published before.

\section{Conclusions}
A few particular exact solutions were derived for two stiff fluids (or massless scalar fields) in non-comoving radial motion in spherical symmetry. The simplest possible physical assumption is investigated when the energy momentum tensors of the fluids are separately conserved. Although that seems too idealized, the same assumption is applied to the different components in $\Lambda$CDM. Spherical symmetry allows for a more complicated thermodynamics though. There could be a temperature gradient along the distinguished radial direction and correspondingly a heat flow. The current paper assumes that the heat conduction of the fluids is negligible hence they can be described as perfect fluids with no heat flow. Two of the presented solutions can be interpreted as stiff fluids with a non-negative energy density over the whole allowed range of coordinates. The other solutions can probably be matched as parts of bigger piece-wise solutions. Due to the separation of variables anzats (16), all solutions considered in this paper have a timelike singularity at the origin of $r$ (except the trivial FRW one) where $A(r)$ vanishes and and a spacelike (cosmological) singularity at the origin of time where $F(t)$ vanishes. Moreover, the solutions are not asymptotically flat (except the static one with k=0).

\section{Acknowledgements}
This work was financially supported by the Department of Energy. The GR calculations were done with GRTensorII, a freely downloadable package for Maple.

\newpage %Just because of unusual number of tables stacked at end
\bibliography{twostiff}% Produces the bibliography via BibTeX.

\begin{thebibliography}{11}
\expandafter\ifx\csname natexlab\endcsname\relax\def\natexlab#1{#1}\fi
\expandafter\ifx\csname bibnamefont\endcsname\relax
  \def\bibnamefont#1{#1}\fi
\expandafter\ifx\csname bibfnamefont\endcsname\relax
  \def\bibfnamefont#1{#1}\fi
\expandafter\ifx\csname citenamefont\endcsname\relax
  \def\citenamefont#1{#1}\fi
\expandafter\ifx\csname url\endcsname\relax
  \def\url#1{\texttt{#1}}\fi
\expandafter\ifx\csname urlprefix\endcsname\relax\def\urlprefix{URL }\fi
\providecommand{\bibinfo}[2]{#2}
\providecommand{\eprint}[2][]{\url{#2}}

\bibitem[{\citenamefont{Letelier and Machado}(1981)}]{let}
\bibinfo{author}{\bibfnamefont{P.~S.} \bibnamefont{Letelier}} \bibnamefont{and}
  \bibinfo{author}{\bibfnamefont{R.}~\bibnamefont{Machado}},
  \bibinfo{journal}{J. Math. Phys.} \textbf{\bibinfo{volume}{22}},
  \bibinfo{pages}{827} (\bibinfo{year}{1981}).

\bibitem[{\citenamefont{Ferrando et~al.}(1989)\citenamefont{Ferrando, Morales,
  and Portilla}}]{fer}
\bibinfo{author}{\bibfnamefont{J.}~\bibnamefont{Ferrando}},
  \bibinfo{author}{\bibfnamefont{J.~A.} \bibnamefont{Morales}},
  \bibnamefont{and} \bibinfo{author}{\bibfnamefont{M.}~\bibnamefont{Portilla}},
  \bibinfo{journal}{Phys.\ Rev. D} \textbf{\bibinfo{volume}{40}},
  \bibinfo{pages}{1027} (\bibinfo{year}{1989}).

\bibitem[{\citenamefont{Haager}(1997)}]{ha1}
\bibinfo{author}{\bibfnamefont{G.}~\bibnamefont{Haager}},
  \bibinfo{journal}{Class. Quantum Grav.} \textbf{\bibinfo{volume}{14}},
  \bibinfo{pages}{2219} (\bibinfo{year}{1997}).

\bibitem[{\citenamefont{Haager}(1998)}]{ha2}
\bibinfo{author}{\bibfnamefont{G.}~\bibnamefont{Haager}},
  \bibinfo{journal}{Class. Quantum Grav.} \textbf{\bibinfo{volume}{15}},
  \bibinfo{pages}{3669} (\bibinfo{year}{1998}).

\bibitem[{\citenamefont{Bayin}(1982)}]{bay}
\bibinfo{author}{\bibfnamefont{S.~S.} \bibnamefont{Bayin}},
  \bibinfo{journal}{Phys. Rev. D} \textbf{\bibinfo{volume}{26}},
  \bibinfo{pages}{1262} (\bibinfo{year}{1982}).

\bibitem[{\citenamefont{Tabensky and Taub}(1973)}]{tab}
\bibinfo{author}{\bibfnamefont{R.}~\bibnamefont{Tabensky}} \bibnamefont{and}
  \bibinfo{author}{\bibfnamefont{A.~H.} \bibnamefont{Taub}},
  \bibinfo{journal}{Commun. Math. Phys.} \textbf{\bibinfo{volume}{29}},
  \bibinfo{pages}{61} (\bibinfo{year}{1973}).

\bibitem[{\citenamefont{Salim and Sautu}(1999)}]{sal}
\bibinfo{author}{\bibfnamefont{J.~M.} \bibnamefont{Salim}} \bibnamefont{and}
  \bibinfo{author}{\bibfnamefont{S.~L.} \bibnamefont{Sautu}},
  \bibinfo{journal}{Class. Quantum Grav.} \textbf{\bibinfo{volume}{16}},
  \bibinfo{pages}{3281} (\bibinfo{year}{1999}).

\bibitem[{\citenamefont{Stephani}(2004)}]{ste}
\bibinfo{author}{\bibfnamefont{H.}~\bibnamefont{Stephani}},
  \emph{\bibinfo{title}{Relativity: an introduction to special and general
  relativity}} (\bibinfo{publisher}{Cambridge University Press},
  \bibinfo{year}{2004}).

\bibitem[{\citenamefont{Murphy}(1960)}]{mur}
\bibinfo{author}{\bibfnamefont{G.}~\bibnamefont{Murphy}},
  \emph{\bibinfo{title}{Ordinary differential equations and their solutions}}
  (\bibinfo{publisher}{D. Van Nostrand Company, Inc.}, \bibinfo{year}{1960}).

\bibitem[{\citenamefont{Hussain et~al.}(1994)\citenamefont{Hussain, Martinez,
  and Nunez}}]{hus}
\bibinfo{author}{\bibfnamefont{V.}~\bibnamefont{Hussain}},
  \bibinfo{author}{\bibfnamefont{E.}~\bibnamefont{Martinez}}, \bibnamefont{and}
  \bibinfo{author}{\bibfnamefont{D.}~\bibnamefont{Nunez}},
  \bibinfo{journal}{Phys. Rev. D} \textbf{\bibinfo{volume}{50}},
  \bibinfo{pages}{3783} (\bibinfo{year}{1994}).

\bibitem[{\citenamefont{Agnese and Camera}(1985)}]{agn}
\bibinfo{author}{\bibfnamefont{A.}~\bibnamefont{Agnese}} \bibnamefont{and}
  \bibinfo{author}{\bibfnamefont{M.}~\bibnamefont{Camera}},
  \bibinfo{journal}{Phys. Rev. D} \textbf{\bibinfo{volume}{31}},
  \bibinfo{pages}{1280} (\bibinfo{year}{1985}).

\end{thebibliography}

\end{document}